%% file: main.tex
\documentclass[12pt,a4paper]{article}
\usepackage{amsfonts}
\usepackage{amsthm}
\usepackage{amsfonts}
\usepackage{amsmath}
\usepackage{amscd}
\usepackage{bbm}
\usepackage{booktabs}
\usepackage[latin2]{inputenc}
\usepackage{t1enc}
\usepackage[mathscr]{eucal}
\usepackage{indentfirst}
\usepackage{graphicx}
\usepackage{xcolor}
\usepackage{hyperref}
\usepackage{graphics}
\numberwithin{equation}{section}
\usepackage{hyperref}
\usepackage[margin=2.9cm]{geometry}
\usepackage[outdir=./images/]{epstopdf}
\usepackage{algorithm}
\usepackage{algpseudocode}
\usepackage{colortbl}

\algrenewcommand\algorithmicrequire{\textbf{Input:}}
\algrenewcommand\algorithmicensure{\textbf{Output:}}

\providecommand{\keywords}[1]
{
  \small	
  \textbf{\textit{Keywords---}} #1
}

\begin{document}
 
\title{Efficient Systematic Reviews: Literature Filtering with Transformers \& Transfer Learning}

\author{
John Hawkins \\ 
Transitional AI Research Group\\
David Tivey \\ 
Royal Australasian College of Surgeons (RACS) \\
} 

\maketitle

\section{Abstract}

Identifying critical research within the growing body of academic work 
is an intrinsic aspect of conducting quality research. Systematic review processes
used in evidence-based medicine formalise this as a procedure that must 
be followed in a research program. However, it comes with an increasing 
burden in terms of the time required to identify the important articles of
research for a given topic. In this work, we develop a method for building 
a general-purpose filtering system that matches a research question, 
posed as a natural language description of the required content, against a 
candidate set of articles obtained via the application of broad search terms. 
Our results demonstrate that transformer models, pre-trained on biomedical 
literature, and then fine tuned for the specific task, offer a promising 
solution to this problem. The model can remove large volumes of irrelevant 
articles for most research questions. Furthermore, analysis of the specific
research questions in our training data suggest natural avenues for further
improvement.

\keywords{Systematic Review, Literature Screening, Machine Learning, BERT, Transfer Learning, Natural Language Processing}

\section{Introduction}

Systematic Reviews are an essential aspect of evidence-based medicine. They 
ensure that researchers are aware of the most relevant and up-to-date literature 
on a particular question. They are, however, labour intensive, time-consuming, error-prone and costly to
undertake\cite{Dinter2021b}. The expanding volumes 
of literature and growing active research areas are increasing the burden 
of thorough systematic reviews \cite{Bastian2010,Miwa2014}. Furthermore, it has been argued that the semantic complexity of scientific literature can hinder the process of systematic reviews, impeding both efficiency and discovery\cite{Westgate2017}.

These issues are affected by a broader set of problems related to the computational
management of academic literature and the curation of datasets of research outcomes. 
This includes matching references 
between datasets or from bibliographies to canonical records \cite{Wu2017},
and identifying the works written by a particular researcher. The problem of matching
a research question to a set of articles sits within the broader problem of 
matching queries to documents, which is plagued by the issue of algorithmically identifying what is semantically relevant between a query and document pair\cite{Ostendorff2020}. 
The systematic review process is one of the more difficult problems in this domain, 
as it requires algorithms that can anticipate what is relevant to a very 
specific research question.

There have been many efforts to build general software applications that assist 
in the systematic review process. Initial approaches focused on algorithms and 
tools that would improve and refine database literature searches,
including the use of empirical studies into methods for refining search terms 
and strategies \cite{Wilczynski2004,Zdravevski2019,Weier2020}, automated query generation\cite{Scells2020},
and machine learning models that emulated the identification of articles of 
\emph{clinical importance}\cite{Aphinyanaphongs2005}.
Most focus has been on methods to improve the efficiency of the 
initial literature screening, labelling articles as either relevant or irrelevant based on title and abstract alone. 
This includes processes that reduce the human 
workload through question-specific predictive models built on an initial set
of human reviewer labels\cite{Frunza2010,Bannach-Brown2019}, expanded to include other kinds of meta-data\cite{Olorisade2019} and applications that update the prioritisation of abstracts in an iterative feedback process \cite{Jonnalagadda2013,Miwa2014,Kontonatsios2020},
or expansion of the articles for consideration using a similarity space and 
initial human selected abstracts \cite{Pham2021}. Recent machine 
learning studies for automation of systematic review focus on alternative machine
learning approaches to building task/question specific models, including the use
of Convolutional Neural Networks (CNNs) \cite{Dinter2021a}, and Transformer Networks \cite{Aum2021}. Multiple studies are making incremental steps toward more general models that reuse what has been learned from previous systematic reviews in different kinds of transfer learning. This includes learning document embeddings that capture information relevant to literature filtering\cite{Kontonatsios2020}, and clustering candidate article abstracts with the articles from previous literature reviews to identify high-quality candidates\cite{Cawley2020}.

However, there is some question as to whether these tools provide 
any real value to systematic review teams\cite{Gartlehner2019}. 
The tools Abstrackr\cite{Wallace2012} and EPPI-Reviewer\cite{Thomas2010}
can both be trained on custom datasets to filter articles toward a 
subset for human reviewers. Independent evaluation has
indicated that these tools will reduce the burden on human reviewers
by a maximum of 50-60\%, respectively. However, on some research questions, they perform considerably worse\cite{Tsou2020}.
Arguably, the biggest challenge for this approach is the requirement 
for an initial set of labelled data for the specific question that is 
large enough to provide a viable model\cite{Aum2021}.

There are alternative approaches that use specialised, but general-purpose models, 
to assist with the systematic review process. These include techniques that identify
factors like the presence of randomised control trials within a given study\cite{Cohen2015}, or the specific text sections that correspond to 
the elements of the PICO\cite{Richardson1995} framework (population; intervention; comparator; outcomes) \cite{Wallace2016,Kang2019,Jin2020}. 
Multiple factors about a paper can be used to build models to filter 
articles in a systematic review, including the article format, purpose 
and the scientific rigor of the 
content \cite{Ambalavanan2020} or evaluating the quality of the research irrespective of the content\cite{Lokker2023}. There are multiple datasets available for these
tasks\cite{Nye2018,DeYoung2020,Ambalavanan2020}, and multiple modelling approaches applied to this task with a recent focus on the use of 
neural networks from Long Short Term Memory (LSTM) \cite{Jin2018} to 
Transformer models\cite{Schmidt2020,Ambalavanan2020}.
These models can be used either to determine exclusion criteria for filtering 
abstracts or given to a human reviewer for consideration in the review process.

Review articles from the systematic review literature itself have 
concluded that machine learning based filters that reduce the volume of articles needed to 
be reviewed are the most promising line of research in the automation of systematic
reviews \cite{Marshall2019,OMara-Eve2015}.

Text data has been shown to be useful across a large number of biomedical 
applications, including the use of unstructured clinical notes in building 
prognostic models\cite{Seinen2022}. It has been noted recently that text 
processing tools are of increasing importance as we attempt to deal with
'infodemics' of the kind generated by the COVID-19 crisis \cite{Vaghela2021}.
 
Using pure text data models for automating systematic review tasks is also
promising. Many of the approaches outlined previously partially use the 
text data in article titles and abstracts within the models. The 
long-term goal remains to build models that replicate the evaluation 
process conducted by human beings, who ultimately look at an article 
with a research question in mind, and then choose whether to include 
it in the systematic review.

The Transformer model\cite{Vaswani2017} has proven to be an effective tool for building text-based classification systems across a range of domains. Part of the utility comes from the fact that it can be pre-trained on the language from a domain and then used for specific tasks, allowing the previous learning to be 'transferred' and applied. It has been shown that transformer models pre-trained on academic abstracts can be
used to create state-of-the-art classifiers for systematic review systems, but only when they train the models on data for a specific task \cite{Aum2021}. This extends previous work using standard feed-forward neural networks to learn a document embedding
that can be employed via transfer learning for a specific systematic review task\cite{Kontonatsios2020}.
The requirement for task-specific training places a burden on teams in 
terms of the time and human effort required to provide labels for the 
initial classifier, that is then iteratively improved with more manual labels. The open research question is whether we can use modern natural language processing transformers to develop general-purpose filters that learn the general correspondence between the phrasing of a research question and the relevant articles.

This research outlines a dataset and comparison of methods for a
general-purpose systematic review model. A single model is trained that 
is given input data combining a research question and information about 
a specific article. It then produces a score 
indicating the likelihood that the specific article would be included by a human reviewer during a literature pre-screen. Our results demonstrate that using BioBERT\cite{Lee2020} version of the BERT Transformer\cite{Devlin2019} pre-trained on biomedical literature can contribute a promising solution to this problem, but the ideal approach depends on the intended process for using a model to improve the systematic review process. We discuss the observed relationships between research questions and machine learning approaches, and conclude that more work is needed to determine the ideal model for all systematic reviews.

\section{Methodology}

\begin{figure}[ht]
\includegraphics[width=1.0\columnwidth]{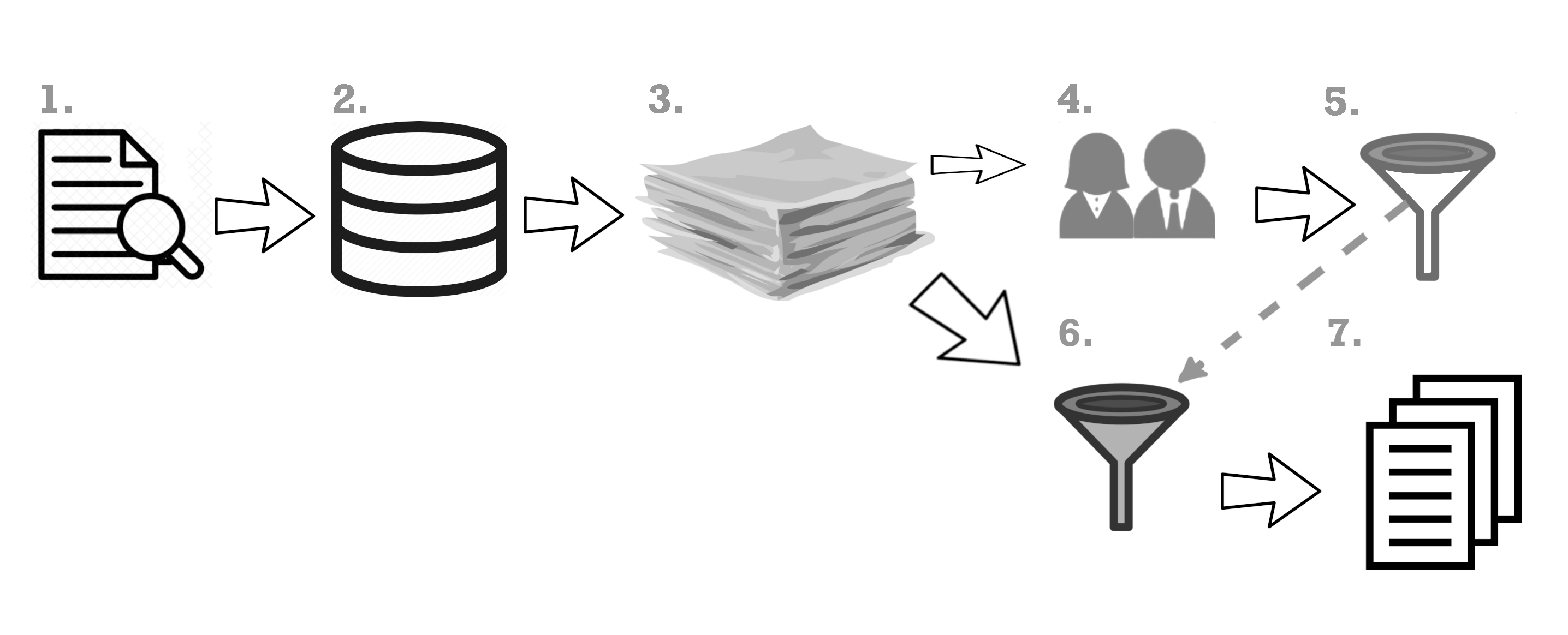}
\caption{Schematic representation of the standard machine learning
approach to building a systematic review filter tool. 
1. Research question is proposed, 
2. Database search is conducted, 
3. List of candidate articles is compiled, 
4. Human annotators review a small selection of articles, 
5. Small annotated collection is used to train a machine learning filter,
6. The machine learning is applied to the remaining documents,
7. Producing the final set for complete review by human experts.}
\label{fig:standard}
\end{figure}

We apply machine learning to develop a process for filtering abstracts 
in a systematic review that is generally applicable and independent of researcher input to provide a question specific label set. 
This means that it takes information about the specific 
research question and the content of all candidate abstracts to 
determine which articles are relevant. This is contrasted with the 
standard approach to building systematic 
review tools with machine learning shown in Figure \ref{fig:standard}.

\begin{figure}[ht]
\includegraphics[width=1.0\columnwidth]{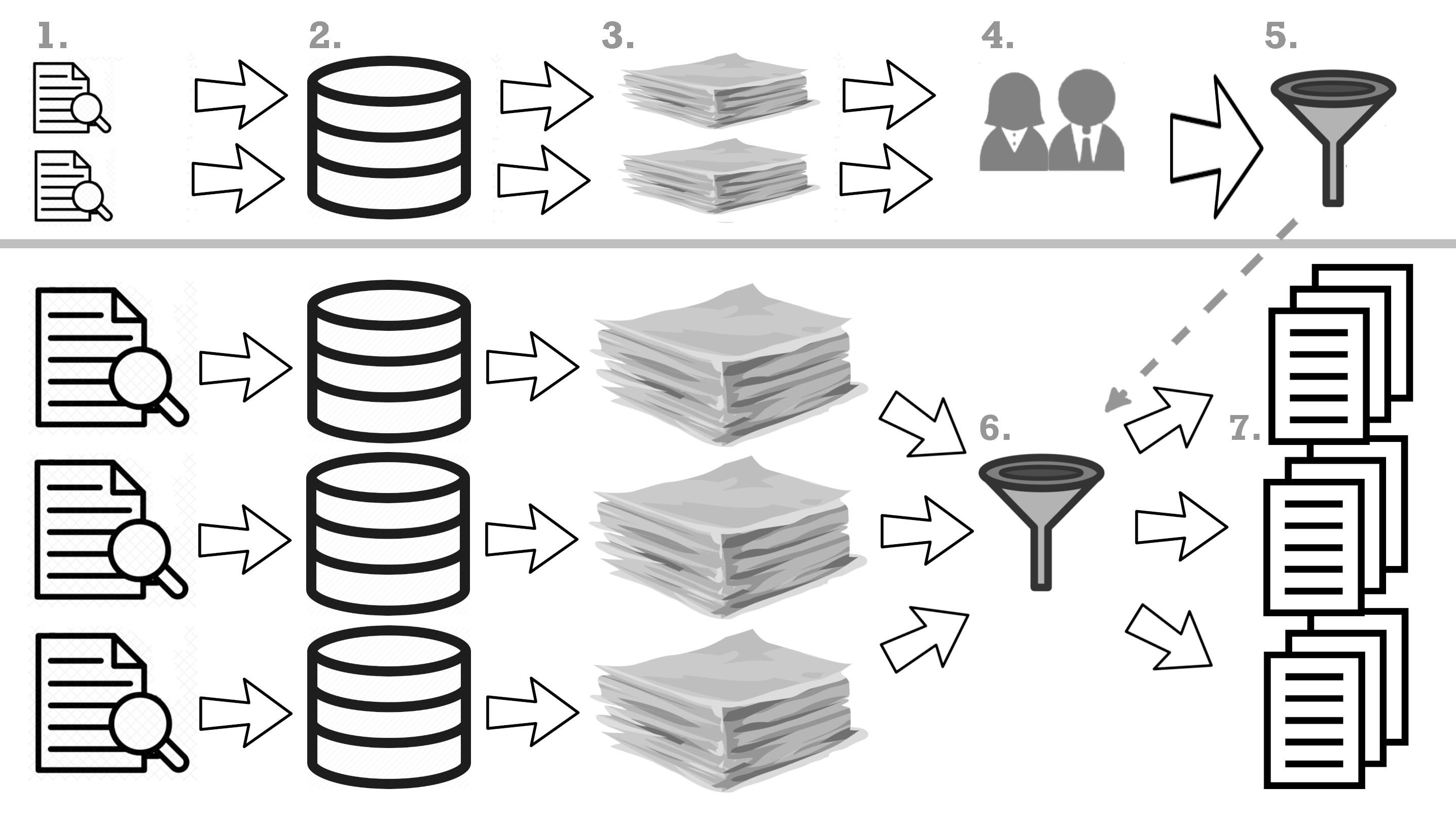}
\caption{Schematic representation of the general systematic review filter process.
The first five steps define how the general filter is trained, the last two the application to new research questions.
1. A set of training research question are taken, 
2. Database searches are conducted for each,
3. Lists of candidate articles are compiled, 
4. Human annotators review these training questions to create an annotated set, 
5. The general filter is trained.
6. The machine learning is applied to filter documents for a new query,
7. Producing the final set for complete review by human experts.}
\label{fig:sysreview}
\end{figure}

Our alternative approach is depicted schematically in Figure \ref{fig:sysreview}. In the top part of the schematic we see that the 
model is trained on data derived from multiple research questions 
that generated human-annotated lists of key 
articles to review. The bottom frame demonstrates that this general purpose 
filter can now be applied to future research questions without the need for human annotators to provide 
an initial seed set of labeled articles for the new research questions. The removal of task-specific annotation for each new research
question is the goal of our research; we use machine learning techniques
that can understand the match between an outline of a research question 
and the content of an article's abstract.

To build this general-purpose filter, we require a dataset of multiple 
research questions, with all of the articles returned from the database searches, as well as the labels resulting from the initial human screening. The curation of this dataset is described below.

\subsection{Dataset Curation}

\input{results/datasets.tex}

For the purposes of data curation, we restrict our initial analysis to research questions that are all  within the general domain of \textit{Therapeutics}. This is a constrained 
domain within the scope of common medical research topics. 
We identified 10 research questions from the ASERNIP-S systematic review archive for which a team of human researchers had manually labelled 
thousands of articles as either \emph{relevant} or \emph{not relevant}. These 
initial questions are summarised in Table \ref{tab:datasets}, showing 
the total number of abstracts retrieved from bibliographic database (e.g, PubMed) searches and 
the number deemed relevant by the human reviewers.

Each of these therapeutics questions has a corresponding research 
question file, which 
is a plain text description of the research program. Each research 
question was derived from the documentation that formed the source 
for both the construction of the database query to extract the candidate
articles and the guidelines by which the human reviewers identified
relevant research.

In addition to the natural language expression of the overall research question, we also generated natural language expressions for each of the four PICO elements independently. The PICO framework has been shown to be well suited to representing therapy questions\cite{Huang2006}, and we conduct
experiments to determine whether text based machine learning tools are 
able to exploit this alternative framing of the research question.

The total dataset contains $34,725$ records with an overall inclusion
rate of $11\%$. However, as shown in Table \ref{tab:datasets} the inclusion rate varies significantly between the research questions, as does the 
volume of articles returned from the database searches.


\subsection{Machine Learning Models}

\begin{table}
\scriptsize
\centering
\caption{Text Match Feature Functions}
\label{tab:features}
\begin{tabular}{l|l|l}
\toprule
Feature Function            &Description                                    &Library \\
\midrule
Length     &Length of text block in chars                          &-   \\
WC         &Count of words in text block                           &-   \\
WL         &Length of words in text (max and mean)                 &-   \\
CWD        &Proportion of words that are not stop words            &-   \\
Match Prop &Proportion of question key words in article text       &-   \\
JD         &Jaro distance between question and article text        &Jellyfish \\
LD         &Levenshtein distance between question and article text  &Jellyfish \\
JI         &Jaccard index between question and article text   &Textdistance \\
SD         &Sorensen distance between question and article text &Textdistance \\
TFIDF      &Term Frequency Inverse Document Frequency       &Scikit-Learn \\
\bottomrule
\end{tabular}
\end{table}

We approach this problem using four general machine learning strategies. The first strategy
is a baseline approach using a set of standard feature engineering approaches and general-purpose 
machine learning algorithms ranging from simple to complex. 
In this approach the research question is contained 
in a single text field of the data and compares with the article's title and abstract using 
text matching algorithms. Each of these text matching approaches generates a numeric value which either describes the context of the research article (title and abstract) or 
indicates the similarity of the text in the question and the article (title and abstract). 
The text features and matching approaches
are drawn from multiple python packages and compiled into the {\it texturizer} python package \cite{texturizer}; the full set of feature generating functions we use are summarised in Table \ref{tab:features}, similarly the full set of standard
machine learning algorithms are summarised in Table \ref{tab:ml}. Note, that the TFIDF approach is only used in some of the experiments, and in all cases this is indicated in the descriptions of the results.

In the second 
approach, we extend the baseline models to include the research question text posed as 
the four independent PICO framework components. Each of the PICO components are matched to the article 
title and abstract independently. The effect of which is to quadruple the text match features, but to 
give the model an indication of what the specific match is between research question and article. For example,
one article might have a high match on population and intervention, whereas another might have a high
match of intervention and outcome.

\begin{table}
\scriptsize
\centering
\caption{Machine Learning Models}
\label{tab:ml}
\begin{tabular}{l|l|l}
\toprule
Model                      &Description                    &Library \\
\midrule
Linear                     &Logistic Regression              &sklearn.linear\_model.LogisticRegression  \\
Naive Bayes (GNB)          &Guassian Naive Bayes             &sklearn.naive\_bayes.GaussianNB   \\
Naive Bayes (CNB)          &Complement Naive Bayes           &sklearn.naive\_bayes.ComplementNB  \\
SVM Classifier             &Support Vector Classifier        &sklearn.svm.LinearSVC   \\
ExtraTrees                 &ExtraTrees Random Forrest        &sklearn.ensemble.ExtraTreesClassifier   \\
LGBM                       &Light Gradient-Boosting Machine  &lightgbm.LGBMClassifier   \\
BERT                       &Pre-Trained BERT Transformer  &transformers.AutoModelForSequenceClassification \\
\bottomrule
\end{tabular}
\end{table}

In the third approach, we employ a set of pre-trained BERT Transformer 
models \cite{Devlin2019} to generate sentence similarity scores as new features. We continue to use the standard text match
features from strategies one and two, but simply append the BERT similarity features to the dataset.
This approach is done using both the standard natural language representation of the research question
and the PICO framework text. We use two general versions of BERT, the BioBERT model\cite{Lee2020} and the BlueBERT model\cite{peng2019}. Summary details for these pre-trained BERT models is shown in Table \ref{tab:transformers},
including the number of heads, layers and the vocabulary size of each model. These parameters provide
an approximation to the complexity of each model.

In the final approach, we fine-tune the pre-trained BERT
Transformer models to directly predict whether an article should be included for full-text review by researchers. In this final approach, no other features are being used, the BERT model is fed the research
question natural language text, followed by the text block end token and then the article's title and
abstract concatenated together. This input representation involves reuse of one of the pre-training
pathways for BERT models in which they are fed two sentences and trained to predict if the sentences have 
the same meaning. In our application, the BERT model is trained to predict if there is sufficient match between the research
question and the article to warrant a full-text review.

In the final fine-tuned transformer approach, we leverage a pre-trained transformer 
language model to process the data as a question-and-answer model. In this 
framing, the research question poses a question, the paper abstract provides the context, and the 
target is the model's yes or no response. We used the native BERT architecture 
to accept two input sentences, typically used in pre-training processes with a 
target value determining if the sentences possess the same meaning. In our usage, this 
pathway indicates if the abstract (sentence 2) is appropriate for the question (sentence 1).

\begin{table}
\centering
\caption{Pre-Trained Transformers}
\label{tab:transformers}
\begin{tabular}{l|r|r|r|l}
\toprule
Transformer Model          &Heads       &Layers  &Vocab Size  &Tuning Data \\
\midrule
BioBERT                    &12          &12      &28996     &PubMed  \\
BlueBERT                   &12          &12      &30522     &PubMed + MIMIC   \\
BlueBERTxL                 &16          &24      &30522     &PubMed + MIMIC   \\
\bottomrule
\end{tabular}
\end{table}

\subsection{Experimental Design}
 
We aim to design a system that can identify relevant abstracts based on a
new research question proposed for systematic review. 
The ideal system cannot rely on a bootstrap set 
of initial labelled examples for the specific research question. 
To align our testing protocol with this goal we use a variation of 
the cross-validation methodology in which each 
fold of the cross-validation consists of all data for one research question. 
This Leave-One-Question-Out (LOQO) provides a close approximation to 
real-world usage, we evaluate how well the model can identify the 
relevant articles based purely on the correspondence between the 
content of the research question and the content of the article's title and abstract.

Note that our approach is far more stringent than the standard cross validation
methodology. Standard cross validation would involve training and testing on
data points from the same research question pool, as such it does not evaluate
performance on truly novel research questions.

\subsection{Evaluation Metrics}

To evaluate our results we use multiple different metrics. The Area Under the Curve ($AUC$)\cite{Hanley1982} provides a value bounded between zero and one that captures the overall ability of a model to rank all articles. It is calculated by iterating over all possible output thresholds for classification and thereby indicates the quality of the model independently of the classification boundary\cite{Melo2013}. 

In addition, we use two metrics to evaluate the performance of the model at the upper and lower end of the article ranking, each of which cater to specific use cases of the models. We first consider the use case of automatically filtering out articles that are not relevant. Eliminating candidate articles is the dominant method by which machine learning filtering tools can provide efficiency for systematic review processes\cite{Marshall2019,OMara-Eve2015}. To evaluate models for this application
we calculate the accuracy in the bottom 50\% ($AccBot50\%$). This means we take the 50\% of records for a given research question that have the lowest scores. We then calculate the percentage of them that were annotated as irrelevant by the human reviewers. The calculation is shown in Equation \ref{eq:bot50}:

\begin{equation}
AccBot50\% = 
\frac
{\sum_{i=0}^{n} \mathbbm{1}_{\{f(Q_i) \, < \, \mathcal{M} \: \& \: l(Q_i) \, = \, 0 \}}}
{\sum_{i=0}^{n} \mathbbm{1}_{\{f(Q_i) \, < \,  \mathcal{M}\}}}
\label{eq:bot50}
\end{equation}

Where $n$ is the number of articles returned for question $Q$. The function $f(Q_i)$ returns the model score for article $i$ of question $Q$, and the function $l(Q_i)$ returns the label. 

We consider a second use case in which a machine learning filter could be used to identify high priority articles to automatically include for full review. To evaluate models for this use case we calculate the accuracy in the top 50 ($AccTop50$). For this metric we take the top 50 highest scoring articles (not percentage) and calculate the accuracy of this set under the assumption that they should all be scored as $relevant$. The calculation is shown in Equation \ref{eq:top50}:

\begin{equation}
AccTop50 = 
\frac
{\sum_{i=0}^{n} \mathbbm{1}_{\{\mathcal{R}(f(Q_i)) \, \le \, 50 \: \& \: l(Q_i) \, = \, 1 \}}}
{\sum_{i=0}^{n} \mathbbm{1}_{\{\mathcal{R}(f(Q_i)) \, \le \, 50 \}}}
\label{eq:top50}
\end{equation}

Where we add an additional function $\mathcal{R}(f(Q_i))$ which returns the rank position of the model scored using the function $f$ on article $i$ of question $Q$. This allows us to calculate the score on only the top 50 ranked articles.

\section{Results}

We show the results for both standard machine learning approaches and transformer models in Tables \ref{tab:aucresults}, Table \ref{tab:botresults} and Table \ref{tab:top50}.
These table display the $AUC$, $AccBot50\%$ and $AccTop50$, respectively. Each modelling approach is
displayed as a row with the results for each of the research questions in the corresponding column. We have highlighted in yellow the best (or equal best) score for each research question.

\begin{table}[h!]
\tiny
 \begin{center}
   \caption{AUC by Model and Research Question}
   \label{tab:aucresults}
   \begin{tabular}{l|r|r|r|r|r|r|r|r|r|r} 
   \hline
    	 experiment	& CARGEL	& CIDP	& EMR	& ESG	& IORT	& LANB	& LMTA	& PBRT	& VERT	& PID\\
   \hline
      	1 - Linear	            &0.436	&0.438	&0.52	&0.612	&0.533	&0.536	&0.57	&0.409	&0.503	&0.506\\
            1 - Naive Bayes (GNB)	&0.694	&0.611	&0.696	&0.5	&0.78	&0.526	&0.614	&0.678	&0.639	&0.745\\
      	1 - Naive Bayes (CNB)	&0.584	&0.587	&0.625	&0.659	&0.605	&0.52	&0.524	&0.563	&0.646	&0.52\\
      	1 - SVM Classifier   	&0.665	&0.493	&0.646	&0.613	&0.594	&0.549	&0.631	&0.583	&0.591	&0.683\\
      	1 - ExtaTrees        	&0.734	&0.568	&0.654	&0.611	&0.758	&0.503	&0.573	&0.644	&0.581	&0.637\\
      	1 - TFIDF-ExtaTrees 	&0.829	&0.769	&0.77	&0.674	&0.748	&0.535	&0.614	&0.696	&0.591	&0.837\\
      	1 - LGBM-TFID        	&0.834	&0.819	&0.795	&0.685	&0.812	&0.537	&0.658	&0.697	&0.555	&0.841\\
    \hline
            2 - Linear-PICO	&0.505	&0.452	&0.402	&0.507	&0.502	&0.515	&0.56	&0.393	&0.466	&0.725\\
      	2 - ExtaTrees-PICO	&0.659	&0.621	&0.742	&0.606	&0.712	&0.499	&0.596	&0.671	&0.616	&0.763\\
            2 - TFIDF-ExtaTrees-PICO	&0.847	&0.755	&0.75	&0.634	&0.74	&0.529	&0.591 &0.699	&0.62	&0.853\\
            2 - LGBM-TFID-PICO	&0.857	&0.764	&0.798	&0.694	&0.721	&0.562	&0.653	&\cellcolor{yellow}0.706	&0.598	&\cellcolor{yellow}0.855\\
    \hline
      	3 - LGBM-BioBERT	&0.841	&0.818	&0.805	&0.696	&0.811	&0.535	&0.67	&0.696	&0.554	&0.831\\
      	3 - LGBM-BlueBERT	&0.845	&0.816	&0.792	&0.689	&\cellcolor{yellow}0.824	&0.539	&0.647	&0.687	&0.556	&0.838\\
      	3 - LGBM-BlueBERTxL	&0.832	&\cellcolor{yellow}0.82	&\cellcolor{yellow}0.809	&0.677	&0.812	&0.547	&0.657	&0.703	&0.575	&0.835\\
      	3 - LGBM-BlueBERT-PICO	&0.865	&0.76	&0.794	&0.702	&0.735	&0.568	&0.63	&\cellcolor{yellow}0.706	&0.578	&0.852\\
    \hline
      	4 - Fine-tune-01e-BioBERT	&0.814	&0.496	&0.502	&\cellcolor{yellow}0.806	&0.808	&\cellcolor{yellow}0.79	&\cellcolor{yellow}0.79	&0.501	&\cellcolor{yellow}0.81	&0.498\\
      	4 - Fine-tune-02e-BioBERT	&\cellcolor{yellow}0.873	&0.5	&0.503	&0.497	&0.626	&0.532	&0.777	&0.502	&0.505	&0.685\\
    \end{tabular}
  \end{center}
\end{table} 

 The AUC describes the ability of the model to rank all articles for the given research question, a value of 
 one would indicate perfect ranking. In some instances we see values very close to 0.5 in Table \ref{tab:aucresults} indicating that in some 
 instances the models are performing no better than chance. Note that all scores were generated using the LOQO process 
 to show the performance of truly novel questions and a list of candidate articles.

\begin{table}[h!]
\tiny
 \begin{center}
   \caption{Bottom 50\% Accuracy by Model and Research Question}
   \label{tab:botresults}
   \begin{tabular}{l|r|r|r|r|r|r|r|r|r|r} 
   \hline
    	 experiment	       & CARGEL	& CIDP	& EMR	& ESG	& IORT	& LANB	& LMTA	& PBRT	& VERT	& PID\\
   \hline
      1 - Linear	           &0.977	&0.855	&0.842	&0.944	&0.956	&0.619	&0.885	&0.692	&0.718	&0.915\\
      1 - Naive Bayes	(GNB)  &0.99	&0.903	&0.916	&0.848	&0.988	&0.608	&0.889	&0.853	&0.808	&0.964\\
      1 - Naive Bayes (CNB)  &0.991	&0.899	&0.892	&0.956	&0.96	&0.674	&0.832	&0.571	&0.824	&0.912\\
      1 - SVM Classifier     &0.99	&0.872	&0.897	&0.945	&0.962	&0.619	&0.907	&0.808	&0.778	&0.951\\
      1 - ExtaTrees	       &0.993	&0.893	&0.902	&0.945	&0.983	&0.587	&0.867	&0.827	&0.772	&0.945\\
      1 - TFIDF-ExtaTrees    &0.996	&0.952	&0.945	&0.955	&0.991	&0.616	&0.898	&0.862	&0.78	&0.981\\
      1 - TFIDF-LGBM         &\cellcolor{yellow}0.997	&\cellcolor{yellow}0.97	&0.954	&0.959	&\cellcolor{yellow}0.994	&0.611	&0.912	&0.849	&0.761	&0.981\\
    \hline
        2 - Linear-PICO	&0.982	&0.857	&0.794	&0.927	&0.953	&0.598	&0.885	&0.696	&0.706	&0.966\\
        2 - ExtaTrees-PICO	&0.99	&0.907	&0.931	&0.944	&0.982	&0.595	&0.894	&0.837	&0.796	&0.968\\
        2 - TFIDF-ExtaTrees-PICO	&\cellcolor{yellow}0.997	&0.947	&0.938	&0.951	&0.987	&0.606	&0.889	&0.853	&0.803	&\cellcolor{yellow}0.984\\
        2 - TFIDF-LGBM-PICO	&\cellcolor{yellow}0.997	&0.955	&0.951	&0.96	&0.978	&0.621	&0.907	&0.849	&0.786	&\cellcolor{yellow}0.984\\
    \hline
      3 - LGBM-BioBERT	&0.994	&0.965	&0.955	&0.962	&0.993	&0.621	&0.92	&0.843	&0.768	&0.978\\
      3 - LGBM-BlueBERT	&\cellcolor{yellow}0.997	&0.965	&0.952	&0.96	&0.993	&0.619	&0.907	&0.827	&0.759	&0.982\\
      3 - LGBM-BlueBERTxL	&0.996	&\cellcolor{yellow}0.97	&\cellcolor{yellow}0.957	&0.959	&\cellcolor{yellow}0.994	&0.621	&0.912	&0.853	&0.774	&0.979\\
      3 - LGBM-BlueBERT-PICO	&\cellcolor{yellow}0.997	&0.95	&0.949	&0.961	&0.98	&0.627	&0.907	&0.862	&0.783	&\cellcolor{yellow}0.984\\
    \hline
      4 - Fine-tune-01e-BioBERT	&0.97	&0.884	&0.886	&\cellcolor{yellow}0.971	&0.963	&\cellcolor{yellow}0.958	&\cellcolor{yellow}0.952	&\cellcolor{yellow}0.887	&\cellcolor{yellow}0.964	&0.887\\
      4 - Fine-tune-02e-BioBERT	&0.982	&0.887	&0.886	&0.885	&0.912	&0.891	&0.948	&\cellcolor{yellow}0.887	&0.888	&0.955\\
    \end{tabular}
  \end{center}
\end{table}

\begin{table}[h!]
\tiny
 \begin{center}
   \caption{Accuracy of Top 50 Predictions by Model and Research Question}
   \label{tab:top50}
   \begin{tabular}{l|r|r|r|r|r|r|r|r|r|r} 
   \hline
    	 experiment	           & CARGEL	& CIDP	& EMR	& ESG	& IORT	& LANB	& LMTA	& PBRT	& VERT	& PID\\
   \hline
      	1 - Linear Baseline	     &\cellcolor{yellow}0.04	&0.14	&\cellcolor{yellow}0.08	&0.04	&0.04	&0.18	&0.22	&0.08	&0.06	&0.02\\
            1 - Naive Bayes	(GNB)    &0.02	&0.12	&0.1	&0.0	&0.0	&0.3	&0.18	&0.08	&0.02	&\cellcolor{yellow}0.08\\
      	1 - Naive Bayes (CNB)	 &0.04	&0.1	&0.1	&0.04	&0.0	&\cellcolor{yellow}0.34	&\cellcolor{yellow}0.24	&0.02	&0.04	&0.04\\
      	1 - SVM Classifier	     &0.0	&\cellcolor{yellow}0.26	&0.06	&0.06	&0.0	&0.26	&0.12	&0.14	&0.02	&0.02\\
      	1 - ExtaTrees	         &0.02	&0.06	&0.04	&0.06	&0.0	&0.3	&0.14	&0.06	&0.14	&0.02\\
      	1 - TFIDF-ExtaTrees	     &0.0	&0.04	&0.0	&0.04	&0.0	&0.28	&0.18	&0.0	&0.16	&0.0\\
      	1 - TFIDF-LGBM	         &0.0	&0.0	&0.0	&0.0	&0.0	&\cellcolor{yellow}0.34	&0.14	&0.0	&\cellcolor{yellow}0.42	&0.0\\
           \hline
      	2 - Linear-PICO	         &\cellcolor{yellow}0.04	&0.16	&0.04	&\cellcolor{yellow}0.14	&0.04	&0.28	&0.16	&\cellcolor{yellow}0.3	&0.02	&0.04\\
      	2 - ExtaTrees-PICO	     &0.02	&0.02	&0.06	&0.02	&0.0	&0.3	&0.14	&0.06	&0.14	&0.02\\
      	2 - TFID-ExtaTrees-PICO	 &0.0	&0.02	&0.0	&0.04	&0.0	&\cellcolor{yellow}0.34	&0.16	&0.04	&0.08	&0.0\\
      	2 - TFID-LGBM-PICO	     &0.0	&0.0	&0.0	&0.0	&0.0	&0.22	&0.1	&0.04	&0.38	&0.0\\
           \hline
      	3 - LGBM-BioBERT 	     &0.0	&0.0	&0.0	&0.02	&0.0	&\cellcolor{yellow}0.34	&0.12	&0.0	&0.34	&0.0\\
      	3 - LGBM-BlueBERT 	     &0.0	&0.0	&0.0	&0.0	&0.0	&0.3	&0.1	&0.02	&0.36	&0.0\\
      	3 - LGBM-BlueBERTxL 	 &0.0	&0.0	&0.0	&0.02	&0.0	&0.3	&0.12	&0.0	&0.3	&0.0\\
      	3 - LGBM-BlueBERT-PICO	 &0.0	&0.0	&0.0	&0.0	&0.0	&0.22	&0.12	&0.04	&0.38	&0.0\\
           \hline
      	4 - Fine-tune-01e-BioBERT	&0.0	&0.12	&0.04	&0.0	&0.0	&0.02	&0.08	&0.1	&0.02	&0.1\\
      	4 - Fine-tune-02e-BioBERT	&0.02	&0.2	&0.06	&0.12	&\cellcolor{yellow}0.12	&0.12	&0.06	&0.1	&0.18	&0.04\\

    \end{tabular}
  \end{center}
\end{table}

Looking at the results in tables \ref{tab:aucresults} and \ref{tab:botresults}, the best-performing models tend to be toward the bottom of the table where we the modelling approaches make use of the pre-trained transformer models.
There does not appear to be a single modelling 
approach that brings us toward optimal performance across all research questions and by extension, all future research questions. However, the fine-tuning approach resulted in improved performance across most research questions. Furthermore, this approach was the only approach to deliver functional AUC for the LANB and VERT questions. All other approaches provided performance that was barely better than random for these questions.

Finally, even within a given category of modelling approach (our 4 experiments), 
there is not a single clear winner in most instances. For example, even though fine-tuning a pre-trained transformer is the strongest overall modelling approach, the performance over one or two epochs of training 
varies considerably between research questions. This makes 
it difficult to determine which
model will perform optimally for a future research question.

\section{Discussion}

This work has conducted a general exploration of machine learning 
approaches to filtering articles returned in a bibliographic database 
search to identify evidence for automated inclusion or exclusion.
We have conducted these experiments using a rigorous Leave-One-Question-Out
methodology, which allows us to understand the ability of each approach 
to perform when presented with truly novel research questions. Our observations from this work 
can be grouped into those associated with machine learning and those 
associated with the formation and phrasing of questions.

\subsection{Machine Learning}

Our work provides substantial evidence that the use of pre-trained transformer models 
offers promise for improving systematic review efficiency through general-purpose machine learning filters. After fine tuning, these models can generalise to recognise articles that are inappropriate for inclusion in the systematic review of genuinely novel research questions. 
When considering the use case of filtering returned articles to exclude the bottom 50\%, the highest accuracy was achieved in 5 of the 10 questions by fine tuning the BioBERT model for a single epoch of training. In the remaining 5 questions, we obtained the highest (or equal highest) accuracy using the pre-trained BERT models to generate similarity match scores between the research questions and article abstract.

However, the ideal methodology for building a general-purpose filter is unclear. For some questions, additional rounds of fine-tuning improved performance, whereas for others, it did not. When the fine-tuned model did not provide optimal accuracy, we found that different BERT models (or methods such as using PICO text) resulted in optimal performance. It remains to be seen whether there is a set of parameters (training rounds and hyper-parameters) or a methodology that will result in an optimal general-purpose filter.

Our comparisons with standard tabular data machine-learning algorithms showed that the BERT-based modelling improved results when considering the AUC or accuracy in the bottom 50\%. However, when we looked at the ability of the models to identify the most likely articles for inclusion, judged by accuracy in the top 50 predictions, we found that simple linear models outperformed all others. The best approach involved using a PICO representation of the research questions and then generating text match features with the PICO elements independently. This surprising result emphasises that the selection of machine learning approaches depends heavily on the model's intended use.

We will continue to explore other approaches to fine-tuning these models, but we have 
also conducted an examination of what made certain research questions easier or harder 
to solve. These insights can help guide the development of tools that might 
satisfy certain subsets of all systematic review questions.

\subsection{Research Question Phrasing}

The ten research questions are shown again in Table \ref{tab:questions} with annotation to define key properties of these inquiries for both the PICO element and standard question formats. These questions were authored by multiple researchers, resulting in various language styles. As such, they reflect a real-world test for automated screening and the flexibility of Natural Language Processing algorithms to identify relevant articles in a dataset extracted from bibliographic databases. Before starting these experiments we hypothesised that questions with greater clinical area coverage would perform better, this was not observed; indicating that the therapy aspect of the research question is more important.

For all standard questions, the Population and Intervention elements of the PICO format were included. Indicating that all questions had the two primary PICO elements that are used in the initial screen for article selection for full-text review. 

Using Microsoft Word, readability statistics were generated for the standard format questions. Except for the LANB and VERT questions, all questions had a Flesch-Kincaid Grade Level (FKGL)\cite{Kincaid1975} above the average level of 15.87 for journal articles, as reported by Kandula and Zeng-Treitlar (2008)\cite{Kandula2008}. As such, the language used by researchers when drafting questions aligns with a journal article corpus. The low FKGL scores may account for the general poor performance of the NLP models, except for Fine-tuned BioBERT, when analysing the LANB and VERT questions. 

In examining the question characteristics and the four questions (PID, CIDP, PBRT, and EMR) that returned the lowest performance in the Fine-tuned BioBERT model (AUC approx 0.5), the incorporation of the Comparator and Outcome PICO components had no substantial impact on the AUC results. Interestingly, two of the four questions with the lowest AUC (Fine-tuned BioBERT) possessed a comprehensive PICO representation (CIDP and EMR), while only the PI elements were present in the PID and PBRT questions. 

Within the PICO elements, the Population, Intervention and Comparator structure vary across research questions. These variations reflect differences in research topics, the health technology being investigated, and the comparator technology considered to be the current clinical practice. In contrast, the structure of the outcome parameter adopted by researchers was a descriptive sentence that included a list of the relevant outcomes. For the PBRT and PID questions, the PICO format in experiment 2 returned the highest AUC (Table \ref{tab:aucresults}). As such, there appears to be some benefit of having separate statements for the PICO elements over the standard question format in these scenarios. This is emphasised by the fact that in experiment 4 these same questions performed very poorly with a fine-tuned BERT model. 

In experiment 4, four questions returned sub-optimal AUC results for the Fine-tuned BioBERT model (PID, CIDP, PBRT and EMR). Two of these, as discussed above, performed well with a PICO representation. The remaining two, performed best using transfer learning with a larger BlueBERT model. This suggests that some research questions contain sufficient complexity (or subtlety) that demands a larger model with a large vocabulary in order to recognise a match between question and article.

Given the importance of the Population and Intervention in the initial screening process, experiments are needed to define the optimal format, level of content and FKGL for the population and Intervention elements in a simple natural language question or the individual PI categories.

\begin{table}
\tiny
\caption{Comparison of Research Question Properties}
\label{tab:questions}
\begin{tabular}{l|llll|llrrr}
\toprule
&\multicolumn{4}{|c|}{PICO Format} &\multicolumn{5}{|c}{Standard Question}\\
\midrule
Question & Population & Intervention & Comparator & Outcome &  PICO & Content & Exclns &  Words &  Flesch-Kincaid GL\\
\midrule
LANB &       B-DS &         N-DS &         SC &   DS-MO &  PI\_O &     Low &        No &     46 &                12.9 \\
PID &      F-DS &         N-MS &         SC &   DS-MO &  PI\_\_ &     Low &         No &     21 &                  25.1 \\
CIDP &     B-DS &            N &    N-MS-EX &   DS-MO &  PICO &     Med &        Yes &    112 &                     19.2 \\
PBRT &         \textasciitilde  &         N-MS &       N-MS &   DS-MO &  PI\_\_ &     Low &       Yes &     29 &                   20.5 \\
IORT &        F-KW &         N-MS &          N &   DS-MO &  PIC\_ &     Med &        No &     72 &                   21.5 \\
ESG &        F-DS &         N-DS &       N-DS &   DS-MO &  PICO &    High &  No &    304 &                24.2 \\
EMR &          F-DS &         N-DS &       N-DS &   DS-MO &  PICO &    High &   No &    150 &                 23.0 \\
CARGEL &   F-KW-MS &            N &         SC &   DS-MO &  PI\_\_ &     Med &       No &     83 &                17.7 \\
LMTA &        F-DS &      N-DS-MS &          N &   DS-MO &  PIC\_ &     Low &    No &     52 &              20.9 \\
VERT &       F-DS &         N-DS &         DS &   DS-MO &  PIC\_ &    High &     Yes &    128 &                   14.7 \\
\bottomrule

\end{tabular}
{\raggedright 
Notes: B=Broad, F=Focused, N=Named, \textasciitilde=Not Defined,
       KW=selected keywords, MS=multi synonyms, 
       DS=descriptive sentence, MO=Multiple Outcomes,
       SC=Standard Care, EX=Exclusions, GL=Grade Level
\par} 
\end{table}

\section{Conclusion}

The transformer machine learning architecture offers unique advantages for building 
natural language processing tools. We have found that pre-trained models can be 
employed to generate text comparison features that outperform a variety of
standard approaches for the majority systematic review questions investigated. 
In addition, we observed that fine tuning one of the models for the specific task 
of identifying suitable articles for a given research question could result in the
best of all surveyed method. However, we also found that there was no single superior 
approach for all research questions. The natural variety in both the phrasing and 
specificity of these research questions prevented determination of a single
ideal modelling approach that could be expected to perform optimally for all 
research questions.

Comparison of the nature of the ten research questions in our study demonstrated that, generally, questions that had substantial detail in the natural language description of the research question performed better. The presence of explicit exclusions in the research question may also pose problems for machine learning systems. The exception to these rules was the EMR question set that had neither exclusions nor a small limited description. The EMR problem required using the larger BERT model to achieve strong performance. The EMR question statement was the second largest in the dataset and contained a complete PICO characterisation, suggesting that research questions with richer detail may require larger pre-trained models.

We have demonstrated that for some research questions using the PICO framework is able to deliver superior performance over simple natural language expressions. This suggests that in some scenarios, the machine learning models benefit from pre-categorising content before generating the classification model's natural language features. This result 
 suggests that custom Transformer architectures (allowing multiple independent text inputs) might provide superior performance.

Overall, judged by AUC, the BERT-based models are consistently superior. When considering a usage process like the automatic exclusion of articles, we look at the accuracy in the bottom 50\% metric. By this account, the BERT models are still generally the strongest performer, although, for certain research questions, they are outperformed by tree-based learning algorithms. However, if we change the usage pattern to consider automatic inclusion of the top 50 predictions, then BERT models perform poorly. For this use case, simple linear models provide the best differentiation.
Whether additional data or fine-tuning can transform the BERT models into consistently superior models for all use cases remains to be seen.

\section*{Acknowledgements}
We thank the systematic review team ASERNIP-S (Royal Australasian College of Surgeons) for providing access to the
historical set of research questions and manually reviewed article lists.

\section*{Declarations}

\subsection*{Competing Interests}

The authors declare that they have no known competing financial interests or personal relationships that could have appeared to influence the work reported in this paper.

\subsection*{Availability of data and materials}

The dataset used for all experiments and the code used
to train, test and analyse all machine learning approaches in this paper will be made available with the publication.

\bibliographystyle{IEEEtran}
\bibliography{refs}

\end{document}

%% file: results/datasets.tex
\begin{table}
\scriptsize
\centering
\caption{Medical Research Questions Datasets}
\label{tab:datasets}
\begin{tabular}{lllrrr}
\toprule
Project &  Clinical Area                &Description & Records & Relevant & Incl Rate \\
\midrule

   LANB &Drug: analgesia          &Post-surgical Nerve Blockade &     766 &314 &41.0\% \\
    PID &Drug: Antibody infusion  &Primary Immunodeficiency &   12269     &1052   &8.6\% \\
   CIDP &Drug: Antibody infusion  &Neurological Inflammation &    4888 &      634 &     13.0\% \\
   PBRT &Radiotherapy             &Partial Breast Radiotherapy &     624 &      154 &24.7\% \\
   IORT &Radiotherapy             &Intraoperative Radiotherapy &    1873 &       93 &5.0\% \\
    ESG &Surgery                  &Endoscopic Sleeve Gastroplasty &    4109 &      314 &7.6\% \\
    EMR &Surgery                  &Endoscopic Mucosal Resection &    3759 &      622 &16.5\% \\
 CARGEL &Surgery                  &Ligament Substitution &    3626 &       66 &1.8\% \\
   LMTA &Surgery                  &Lung Microwave Tissue Ablation &     452 &       68 &15.0\% \\
   VERT &Surgery                  &Oesteoporotic Vertebral Fractures &    2359 &643 &27.3\% \\
\bottomrule
\end{tabular}
\end{table}